\documentclass[aps,preprint,tightenlines,groupedaddress]{revtex4}%
\usepackage{amssymb}
\usepackage{amsfonts}
\usepackage{amsmath}
\usepackage{subfig}
\usepackage{graphicx}%
\providecommand{\U}[1]{\protect\rule{.1in}{.1in}}
\newtheorem{theorem}{Theorem}
\newtheorem{acknowledgement}[theorem]{Acknowledgement}

\begin{document}
\title{Do uniform tangential interfacial stresses affect adhesion?}
\author{Nicola Menga$^{1,3}$, Giuseppe Carbone$^{1,2,3}$, Daniele Dini$^{1,3}$}
\affiliation{$^{1}$Department of Mechanics, Mathematics and Management, Politecnico di
Bari, v.le Japigia 182, 70126 Bari - Italy}
\affiliation{$^{2}$Physics Department M. Merlin, CNR Institute for Photonics and
Nanotechnologies U.O.S. Bari via Amendola 173, 70126 Bari, Italy}
\affiliation{$^{3}$Department of Mechanical Engineering, Imperial College London, South
Kensington Campus, Exhibition Road London SW7 2AZ, United Kingdom}

\begin{abstract}
We present theoretical arguments, based on linear elasticity and
thermodynamics, to show that interfacial tangential stresses in sliding
adhesive contacts does not affet at all the adhesive behavior of the system,
which then follows the classical JKR\ solution. Our finding explains the
experimental observation of Vorvolakos and Chaudhury in 2003 (Ref.
\cite{Vorvolakos2003}), who found that the contact area of a PDMS\ sphere
remains constant during sliding and is in agreement with the JKR solution, at
least up to velocity of 1mm/s, and of Carpick et al. in Ref. \cite{Carpick},
who observed that the friction force between a platinum-coated atomic force
microscope (AFM) tip and the surface of mica in ultrahigh vacuum (UHV) varies
with load in proportion to the contact area predicted by the
Johnson-Kendall-Roberts (JKR). We show that a reduction of the contact area,
experimentlly observed at higher sliding speeds, can be caused by a reduction
of the density of adhesive bonds as the velocity is increased, or caused by
the repulsive energy term associated with the stress spatial fluctuation at
the interface. This may explain why adhesion is completely masked at
relatively large sliding velocities.

\textit{This version of the paper follows the publication of the Corrigendum:}

\textit{Nicola Menga, Giuseppe Carbone, Daniele Dini: Corrigendum to
\textquotedblleft Do uniform tangential interfacial stresses enhance
adhesion?\textquotedblright\ [Journal of the Mechanics and Physics of Solids
112 (2018) 145--156], Journal of the Mechanics and Physics of Solids, 133,
103744, https://doi.org/10.1016/j.jmps.2019.103744, available on line since 8
October 2019.}

\end{abstract}
\maketitle

\section{Introduction}

In the last decades, contact mechanics, and in particular the effect of
physical interactions occurring at the interface between elastic and
viscoelastic solids, has found increasing scientific interest, mostly boosted
by practical applications, such as tires, seals, bio-inspired climbing robots,
and adhesive gloves. The contact behavior of such systems has been studied by
many authors relying on different approaches: analytical techniques
\cite{Hunter, Grosch, Persson2001, Persson2010, kalkerPanek, expwear, menga1},
advanced numerical simulations \cite{Padovan1, Padovan2, Padovan3,
FEMmultitau, FrancesiFEM,mengavisco,mengaadhesive,mengaRLRB} and experimental
investigations \cite{Schapery1969, Faisca2001, Odegard2005, carbonr friction,
creton}.

Among the many factors influencing interactions occurring in contact problems,
the interplay between shear stresses (and associated frictional response) and
adhesion in elastic contacts is a long-standing tribological problem. Many
authors have contributed to shed light on the relation between friction,
adhesion and contact area, motivated by the relevance that this phenomenon has
in a countless number of engineering applications involving \textit{e.g.} wear
\cite{Extrand1991,Theodore1992,MengaCiava,MengaRussi}, shear resistance
\cite{newby1995,newby1998}, tire friction \cite{Persson2004,Pohl1999},
electric resistance \cite{Yi2005}, mixed and boundary lubrication
\cite{Adams2001,Stanley1990}, and slippery prosthetic devices
\cite{Nusbaum1979,Dong2000}. Therefore, both experimental
\cite{Savkoor1977,Homola1990,Vorvolakos2003,Degennes2007} and theoretical
investigations \cite{Savkoor1977,Johnson1997}\ have been carried out on this
specific topic with the aim of providing additional insights into the adhesive
behavior of frictional contacts. In particular, most of them seem to indicate
that the presence of relative sliding and friction at the interface always
leads to a reduction of the contact area, and, therefore weakens the adhesion
strength. This phenomenon, which is known as a friction induced transition
from the adhesive JKR regime \cite{jkr} to the adhesiveless Hertz regime, is
usually explained by relying on the arguments presented firstly by Savkoor and
Briggs \cite{Savkoor1977}, and then by Johnson \cite{Johnson1997}. However, we
cannot ignore the fact that the theoretical arguments presented in Ref.
\cite{Savkoor1977} holds true only for the case of contacts in the presence of
full stick between the meeting surfaces, where the occurrence of slip at the
contact interface is prevented from taking place (\textit{i.e.} in the
presence of uniform tangential displacement). Therefore, this theory is not
well suited to deal with sliding contacts as those addressed in Refs
\cite{Homola1990,Vorvolakos2003,Degennes2007}, where gross slip conditions
between almost perfectly smooth surfaces are investigated, leading to
significantly different conclusions: the presence of slip at moderate
velocities does not lead to any reduction of contact area. Moderate slip
velocities, then, do not hinder adhesion. A loss of adhesion is instead
observed at higher sliding velocities, and is often related to stick-slip
transitions. However, even the loss of adhesion observed at high velocity,
cannot be explained with the no-slip Savkoor and Briggs' theory, where the
presence of the tangential stress singularity at the edge of the contact makes
the energy release rate increase, weakens the adhesive bond and leads to a
decrease of the contact area. On the contrary, in presence of gross slip at
the interface, the tangential stresses singularity is prevented from
occurring, thus impeding the mechanism described by Savkoor and Briggs
\cite{Savkoor1977} from taking place. From the theoretical point of view,
every existing attempt at showing the relationship between shear and adhesion
has overlooked this aspect of the problem (see e.g. \cite{Johnson1997,
persson-err}). In gross slip, a first possible mechanism of adhesion loss has
been described firstly by Schallamach \cite{Schallamach1963} and then by
Chernyak and Leonov \cite{Chernyak1986}, where the loss of adhesion can be
attributed to breaking and partial reformation of adhesive bonds during
sliding. A second mechanism could be, instead, related to the effect of
spatial frictional stress fluctuations and the interface, which as we shown in
the paper, lead to an additional repulsive surface energy term, which may
counterbalance the adhesion forces. Unfortunately, to the best of our
knowledge, experimental evidence aimed at shedding light on this is lacking,
and this is mainly due to the fact that investigations in the regime of
interest require very accurate instrumentation and analytical techniques.

In this study, we focus on adhesive sliding contacts between perfectly smooth
surfaces under the condition that gross slip takes place at moderate
velocities. We treat the exemplar case of a smooth rigid sphere sliding on a
soft elastic half-space, and present a rigorous thermodynamic treatment of the
contact behavior at the interface aimed at deepening the understanding of the
influence of tangential stresses on adhesion and contact area. We have spent
much time investigating this peculiar and yet under-investigated phenomenon,
supported by the theoretical results presented in this contribution and by a
novel interpretation of the few existing experimental results reported in the
literature on this topic.

\section{Formulation\label{formulation}}

\begin{figure}[ptbh]
\begin{center}
\includegraphics[width=0.9\textwidth]{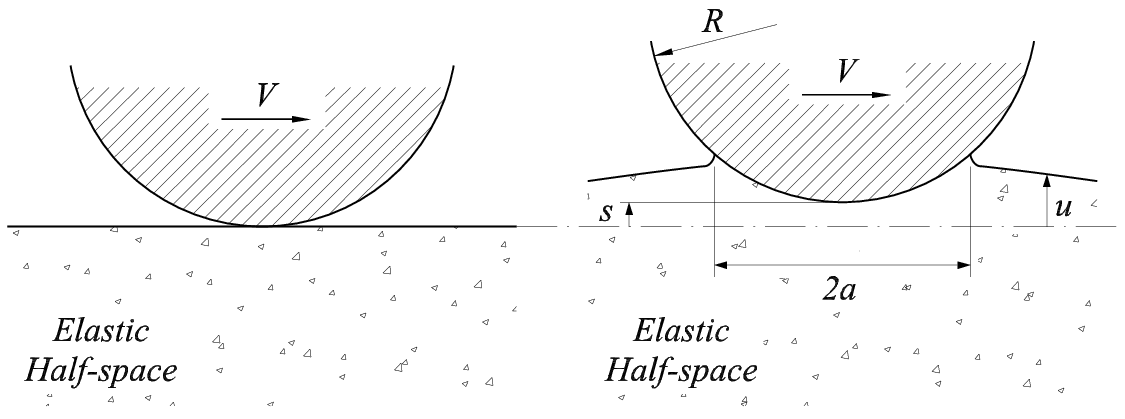}
\end{center}
\caption{The geometrical scheme of the contact problem: a rigid spherical
indenter of radius $R$ is sliding at constant velocity $V$ over an elastic
half-space. In the figure, $a$ is the contact area radius, and $s$ is the
contact separation.}%
\label{fig1}%
\end{figure}

We consider the case of an elastic half-space in sliding contact with a
spherical rigid indenter of radius $R$, as shown in Fig. \ref{fig1}. As a
result of the contact interactions, the half-space is loaded with a certain
distribution of normal and tangential stresses on a portion $\Omega$ of its
surface (namely the contact domain). In particular, we focus on the specific
case where the tangential stress at the interface are uniformly distributed
with value $\tau$. This choice is strictly related to the observation that in
sliding contacts of soft polymeric materials the interfacial tangential
stresses do not follow a Coulomb friction law but rather they are almost
uniform at the interface \cite{Gao2004,Chateauminois2008}.

For the system at hand (see Fig. \ref{fig1}), we define as $u\left(
\mathbf{x}\right)  $ and $v\left(  \mathbf{x}\right)  $ the normal and
tangential displacement fields respectively.

Recalling that the tangential stresses are uniformly distributed
on the contact area, the elastic energy stored in the body can be calculated
as%
\begin{equation}
\mathcal{E}=\frac{1}{2}\int_{\Omega}d^{2}x\sigma\left(  \mathbf{x}\right)
u\left(  \mathbf{x}\right)  +\frac{1}{2}\tau W,\label{Eel_2}%
\end{equation}
where $\Omega$ is the contact domain, and the quantity%
\begin{equation}
W=\int_{\Omega}d^{2}xv\left(  \mathbf{x}\right)  =v_{m}A\label{W}%
\end{equation}
is, what we call, the \textit{displaced tangential volume}, being $v_{m}$ the
average tangential displacement in the contact area. The internal energy of
the system is then given by
\begin{equation}
\mathcal{U}\left(  s,W,A\right)  =\mathcal{E}\left(  s,W,A\right)
-\Delta\gamma A,\label{U}%
\end{equation}
where $\Delta\gamma$ work of adhesion and $A=\left\vert \Omega\right\vert $ is
the contact area. Finding the minimum of $\mathcal{U}\left(  s,W,A\right)  $
allows to find the contact solution when the state parameters are the
separation $s$ (or equivalently the penetration $\delta=-s$), the tangentially
displaced volume $W$, and the contact area $A$. However, in our problem, the
state variables are $\left(  s,\tau,A\right)  $, as the shear stress $\tau$ is
uniform in the contact area and is actually kept constant as the system
configuration changes towards the final equilibrium state. In such a case,
there is a certain amount of mechanical energy associated with the constant
stress field $\tau$. Therefore, we prefer to consider the following potential
energy associated with the uniformly distributed stress $\tau$ that can be
determined by performing the following Legendre transform (see \cite{callen}%
):
\begin{equation}
\mathcal{H}=\mathcal{U-}\left(  \frac{\partial\mathcal{U}}{\partial W}\right)
_{s,A}W=\mathcal{U-}\tau W,\label{legendre transform}%
\end{equation}
where, in fact, the term $-\tau W$ is the potential energy associated with the
uniform stress distribution $\tau$. This leads to define the new thermodynamic
potential%
\begin{equation}
\mathcal{H}\left(  s,\tau,A\right)  =\frac{1}{2}\int_{D}d^{2}x\sigma\left(
\mathbf{x}\right)  u\left(  \mathbf{x}\right)  -\frac{1}{2}\tau W-\Delta\gamma
A.\label{H_2}%
\end{equation}
By changing the contact area of a quantity $\delta A$ at fixed separation $s$
and shear stress $\tau$, we get%
\begin{equation}
\left(  \frac{\partial\mathcal{H}}{\partial a}\right)  _{s,\tau}=-\left(
\frac{\partial}{\partial a}\frac{4\tau^{2}a^{3}}{E^{\ast}}\right)  _{\tau}%
\neq0\label{equil}%
\end{equation}
This can be easily shown by observing that the change $\delta\mathcal{U}$ of
internal energy $\mathcal{U}$ of the system as the contact area $A=\pi a^{2}$
is changed of the quantity of $\delta A=2\pi a\delta a$, while keeping
constant $s$ and $\tau$, is equal to the work $\delta L$ done by the
tangential forces i.e.%
\begin{equation}
\left(  \delta\mathcal{U}\right)  _{s,\tau}=\delta L\label{uno}%
\end{equation}
where%
\begin{equation}
\delta L=\int_{A}\tau\delta v\left(  \mathbf{x}\right)  dA=\tau\int_{A}\delta
v\left(  \mathbf{x}\right)  dA=\tau\delta W-2\pi a\tau\bar{v}\delta
a\label{due}%
\end{equation}
and
\begin{equation}
\bar{v}=\frac{1}{2\pi}\int d\theta v\left(  a\cos\theta,a\sin\theta,a\right)
\label{tre}%
\end{equation}
In Eq. (\ref{due}) we have used that%
\begin{equation}
\delta W=\delta\int_{A}v\left(  \mathbf{x}\right)  dA=\int_{\delta A}v\left(
\mathbf{x}\right)  dA+\int_{A}\delta v\left(  \mathbf{x}\right)  dA=2\pi
a\bar{v}\delta a+\int_{A}\delta v\left(  \mathbf{x}\right)  dA\label{quattro}%
\end{equation}
Moreover, recalling Eq. (A.8) and keeping the assumption of Poisson's ratio
$\nu=0.5$ we get
\begin{equation}
\bar{v}=\frac{6\tau a}{\pi E^{\ast}}\label{v}%
\end{equation}
Now, replacing Eq. (\ref{v}) in Eq. (\ref{due}) and recalling Eq. (\ref{uno})
we obtain Eq. (\ref{equil}) i.e.%
\begin{equation}
\left[  \frac{\partial\left(  \mathcal{U}-\tau W\right)  }{\partial a}\right]
_{s,\tau}=\left(  \frac{\partial\mathcal{H}}{\partial a}\right)  _{s,\tau
}=-12\frac{\tau^{2}a^{2}}{E^{\ast}}=-\left(  \frac{\partial}{\partial a}%
\frac{4\tau^{2}a^{3}}{E^{\ast}}\right)  _{\tau}\neq0\label{lavoro forze2}%
\end{equation}
Equation (\ref{equil}) allows to find the equilibrium solution of the system.
Furthermore, we note that%
\begin{equation}
\left(  \frac{\partial\mathcal{H}}{\partial s}\right)  _{A,\tau}=F,\label{F}%
\end{equation}
where $F$ is the total remote tractive force acting on the sphere.

In what follows we consider the case of soft polymeric materials. We assume
that the material is incompressible (Poisson's ratio $\nu=0.5$), which makes
the tangential and normal elastic fields uncoupled, and that the elastic
substrate is a half-space, in contact with a rigid sphere of radius $R$, over
a circular area of radius $a\ll R$ (see figure \ref{fig1}b). The displaced
tangential volume is $W=\pi a^{2}v_{m}$, where $v_{m}$ can be easily estimated
by observing that tangential strain must be of order $v_{m}/a$. Therefore,
because of linear elasticity, being $G$ the shear modulus, the tangential
stress $\tau$ must be of order $Gv_{m}/a$, which gives%
\begin{equation}
v_{m}=K\frac{\tau}{G}a, \label{stima dello spostamento medio}%
\end{equation}
with $K$ being a constant of order unity. Equation
(\ref{stima dello spostamento medio}) can be also obtained by dimensional
arguments \cite{Buckingham1914,Buckingham1915,Buckingham1915b}. Recalling the
Gibbs phase rule \cite{Gibbs}, there must be a state equation linking the
three quantities $s,\tau,a$. Therefore, at equilibrium, the number of
independent quantities is two, and one can write $v_{m}=f\left(
G,a,\tau\right)  $. Now, choosing as fundamental dimensional quantities the
shear modulus $G$ and the contact radius $a$, following Buckingham's theorem
\cite{Buckingham1914,Buckingham1915,Buckingham1915b}\ we write $v_{m}%
/a=g\left(  \tau/G\right)  $, and because of linear elasticity, we conclude
that the functional form of $g\left(  \tau/G\right)  $ must be a relation of
proportionality. Thus, the reduced displacement $v_{m}/a$ is proportional to
the reduced tangential stress $\tau/G$, leading again to Eq.
(\ref{stima dello spostamento medio}). In appendix \ref{appendice1} we present
an analytical derivation of the interfacial tangential displacement field
caused by uniform tangential tractions applied on a circular area of an
elastic half-space, and demonstrate that $K=2/\pi$. Therefore Eq.
(\ref{stima dello spostamento medio}) becomes%
\begin{equation}
v_{m}=\frac{2}{\pi G}\tau a=\frac{8}{\pi E^{\ast}}\tau a,
\label{real mean value.}%
\end{equation}
where we have introduce the reduced elastic modulus $E^{\ast}=E/\left(
1-\nu^{2}\right)  $. It follows that $W=\pi a^{2}v_{m}=8\tau a^{3}/E^{\ast}$,
and the energy term associated with the uniform stress distribution $\tau$ is
then%
\begin{equation}
\tau W=\frac{8\tau^{2}a^{3}}{E^{\ast}}. \label{mechanical energy}%
\end{equation}
Notably, since $\nu=0.5$, the normal displacement field is uncoupled from the
tangential one. Thus, we can use the solution of the frictionless adhesive
contact between a sphere and a half space. This solution is reported in
\cite{carbonebook}, and it is easy to recover as the superposition of a rigid
flat punch solution and Hertz solution. Therefore, the normal displacement
field in the contact area is%
\[
u\left(  \mathbf{x}\right)  =s+\frac{\left\vert \mathbf{x}\right\vert ^{2}%
}{2R},
\]
and the normal stress field at the interface%
\begin{equation}
\sigma\left(  \mathbf{x}\right)  =\sigma_{0}\left(  1-\frac{\left\vert
\mathbf{x}\right\vert ^{2}}{a^{2}}\right)  ^{-1/2}+\sigma_{1}\left(
1-\frac{\left\vert \mathbf{x}\right\vert ^{2}}{a^{2}}\right)  ^{1/2},
\label{general solution}%
\end{equation}
with%
\begin{align}
\sigma_{0}  &  =\frac{1}{\pi}E^{\ast}\left(  \frac{s}{a}+\frac{a}{R}\right)
\label{p0}\\
\sigma_{1}  &  =-\frac{1}{\pi}E^{\ast}\frac{2a}{R}. \label{p1}%
\end{align}
As expected, when $\sigma_{0}$ is different from zero the stress distribution
has a square root singularity as $\left\vert \mathbf{x}\right\vert \rightarrow
a$. Negative values of $\sigma_{0}$ [\textit{i.e.} $a<\left(  -sR\right)
^{1/2}$] are not physically acceptable since they would cause interpenetration
of solids near the edges of the contact. Therefore only non-negative values of
$\sigma_{0}$ are admissible, which is equivalent to say that the contact
radius $a$ must satisfy the following inequality
\begin{equation}
a\geq a_{Hz}, \label{non interpenetration condition}%
\end{equation}
where $a_{Hz}=\left(  -sR\right)  ^{1/2}$ is the Hertzian contact radius that
would be obtained in case of adhesiveless contacts at a given separation $s$.
Moreover, for $a>a_{Hz}$ the interfacial tractive stress at the edge of the
contact diverges towards infinitely large values. The latter scenario strictly
requires tractive stresses to be developed in the contact, \textit{i.e.} it
can happen only in presence of adhesive forces. In absence of adhesion,
instead, $\sigma_{0}=0$ and $a=a_{Hz}$.

\subsection{Fixed separation}

Now, given the separation $s$, we can calculate the thermodynamic potential
$\mathcal{H}\left(  s,\tau,a\right)  $ from Eq. (\ref{H_2}) as%
\begin{equation}
\mathcal{H}\left(  s,\tau,a\right)  =E^{\ast}s^{2}\left(  a+\frac{2}{3}%
\frac{a^{3}}{Rs}+\frac{1}{5}\frac{a^{5}}{R^{2}s^{2}}\right)  -\frac{4\tau
^{2}a^{3}}{E^{\ast}}-\pi\Delta\gamma a^{2} \label{total energy sphere}%
\end{equation}
Notably, from Eq. (\ref{F}) the normal force is%
\begin{equation}
F=\left(  \frac{\partial\mathcal{H}}{\partial s}\right)  _{A,\tau}=E^{\ast
}\left(  2sa+\frac{2}{3}\frac{a^{3}}{R}\right)  . \label{Normal Force}%
\end{equation}

In dimensionless terms, Eq. (\ref{total energy sphere}) becomes%
\begin{equation}
\mathcal{\tilde{H}}=\tilde{a}-\frac{2}{3}\tilde{a}^{3}+\frac{1}{5}\tilde
{a}^{5}-\tilde{\tau}^{2}\tilde{a}^{3}-\Delta\tilde{\gamma}\tilde{a}^{2},
\label{dimensionless energy}%
\end{equation}
where we have defined the following reduced quantities:%
\begin{equation}
\mathcal{\tilde{H}}=\frac{R^{2}\mathcal{H}}{E^{\ast}a_{Hz}^{5}};~\tilde{\tau
}=\frac{2R\tau}{E^{\ast}a_{Hz}};~\Delta\tilde{\gamma}=\pi\frac{R^{2}%
\Delta\gamma}{E^{\ast}a_{Hz}^{3}};~\tilde{a}=\frac{a}{a_{Hz}}.
\label{reduced quantities}%
\end{equation}
Finally, enforcing Eq. (\ref{equil}) gives%

\begin{equation}
\left(  \frac{\partial\mathcal{\tilde{H}}}{\partial\tilde{a}}\right)
_{\tilde{\tau}}=1-2\tilde{a}^{2}+\tilde{a}^{4}-3\tilde{\tau}^{2}\tilde{a}%
^{2}-2\Delta\tilde{\gamma}\tilde{a}=-3\tilde{\tau}^{2}\tilde{a}^{2},
\label{equilibrium condition}%
\end{equation}
which gives the classical JKR\ solution \cite{jkr}%
\begin{equation}
1-2\tilde{a}^{2}+\tilde{a}^{4}-2\Delta\tilde{\gamma}\tilde{a}=0
\label{classical JKR}%
\end{equation}
and allows to determine the contact radius at equilibrium $\tilde{a}_{eq}$.

\subsection{Fixed Load}

In a similar way it is possible to address the contact case characterized by a
constant normal force, $F$ (configuration commonly adopted in experiments
\cite{Savkoor1977,Homola1990,Vorvolakos2003}). Again, we need to move to
another thermodynamic potential $\mathcal{G}$ by means of the new Legendre
transformation%
\begin{equation}
\mathcal{G}=\mathcal{H-}\left(  \frac{\partial\mathcal{H}}{\partial s}\right)
_{s,A}s=\mathcal{H-}Fs. \label{gibbs energy}%
\end{equation}

Consequently, the equilibrium condition becomes%
\begin{equation}
\left(  \frac{\partial\mathcal{G}}{\partial a}\right)  _{F,\tau}=-\left(
\frac{\partial}{\partial a}\frac{4\tau^{2}a^{3}}{E^{\ast}}\right)  _{\tau}
\label{new equlibrium}%
\end{equation}
Moreover, by using Eqs. (\ref{total energy sphere}, \ref{Normal Force}) we get%
\begin{equation}
s=\frac{F}{2E^{\ast}a}-\frac{a^{2}}{3R}, \label{displacement}%
\end{equation}
and%
\begin{equation}
\mathcal{G}\left(  F,\tau,a\right)  =-\frac{F^{2}}{4E^{\ast}a}+\frac{Fa^{2}%
}{3R}+\frac{4}{45}\frac{E^{\ast}a^{5}}{R^{2}}-\frac{4\tau^{2}a^{3}}{E^{\ast}%
}-\pi\Delta\gamma a^{2}, \label{Gpot}%
\end{equation}
which in dimensionless form becomes%
\begin{equation}
\mathcal{\tilde{G}=}\frac{R^{2}\mathcal{G}}{E^{\ast}a_{Hz}^{5}}=-\frac
{4}{9\tilde{a}}-\frac{4\tilde{a}^{2}}{9}+\frac{4\tilde{a}^{5}}{45}-\tilde
{\tau}^{2}\tilde{a}^{3}-\Delta\tilde{\gamma}\tilde{a}^{2},
\label{dimensionless_G}%
\end{equation}
where the definitions of $\tilde{\tau}$ and $\Delta\tilde{\gamma}$ are given
in Eq. (\ref{reduced quantities}), and $a_{Hz}=-\left(  3FR/4E^{\ast}\right)
^{1/3}$.

Thus, from Eq. (\ref{new equlibrium}) we obtain%
\begin{equation}
\frac{\partial\mathcal{\tilde{G}}}{\partial\tilde{a}}\mathcal{=}\frac
{4}{9\tilde{a}^{2}}-\frac{8\tilde{a}}{9}+\frac{4}{9}\tilde{a}^{4}-3\tilde
{\tau}^{2}\tilde{a}^{2}-2\Delta\tilde{\gamma}\tilde{a}=-3\tilde{\tau}%
^{2}\tilde{a}^{2}, \label{contact area given load}%
\end{equation}
which allows to determine the equilibrium contact area at a given load and
still provide the same classical JKR \cite{jkr} solution as before.

\section{Results and discussion}

The above result shows that \textit{the presence of uniform tangential
stresses at the interface does not modify the effective energy of adhesion,
therefore the contact behavior follows the classical JKR\ solution even in
sliding at uniform frictional stress.} This result is in agreement with the
experiments carried out by Vorvolakos and Chaudhury in 2003 (Ref.
\cite{Vorvolakos2003}), who observed experimentally that the contact area of a
PDMS\ sphere is constant during sliding, at least up to velocity of 1mm/s, and
well predicted by the JKR solution. Our result also explain the experimental
outcomes of Carpick et al. in Ref. \cite{Carpick}, who observed that the
friction force between a platinum-coated atomic force microscope (AFM) tip and
the surface of mica in ultrahigh vacuum (UHV) varies with load in proportion
to the contact area predicted by the Johnson-Kendall-Roberts (JKR).

The above arguments seems to disprove the fact that sliding friction
necessarily hinders adhesion, thus contravening what is commonly accepted in
the scientific literature. To answer this question we need to revisit some of
the most interesting experiments on the topic. Several of them
\cite{Homola1990,Vorvolakos2003,Degennes2007} show that adhesion totally
disappears for sufficiently high sliding velocity. Among them, we identify the
work by Vorvolakos and Chaudhury \cite{Vorvolakos2003} as one of the most
interesting in the field. Focusing on Fig. 7 of the cited paper
\cite{Vorvolakos2003}, we note that the contact area reduction is observed
only for sliding velocities larger than $2$-$3$ mm/s, and the resulting
contact area is even smaller than the Hertzian predicted value. With respect
to this peculiar behavior, most of the existing literature refers to the
seminal paper by Savkoor and Briggs \cite{Savkoor1977}, where the authors
justify the contact area reduction relying on fracture mechanics arguments.
Specifically, they assume uniform tangential displacement at the interface
(i.e. they are controlling the tangential displacement rather then the
tangential frictional tractions), which gives rise to a shear stress
singularity at the contact area edges which, in turn, is modelled as a mode II
crack opening. Eventually, the Griffith equilibrium condition predict a
reduction of the contact area size. However, from a theoretical point of view,
we observe that the main assumption behind this result is that the
displacement distribution over the contact area is uniform. This physical
picture clearly refers to the case of a fully stuck contact under externally
applied tangential load. On the contrary, when gross sliding is developed
between the two interfaces, frictional shear stresses are prescribed instead
of continuity of displacements across the interface. As shown in several
previous work e.g. \cite{Chateauminois2008, scheibert}, in the case of soft
polymers shear stresses are distributed more or less uniformly (apart from
fluctuations, also covered in this study) over the sliding contact area. We
therefore conclude that since under gross slip there is no shear stress
singularity in the contact, the model proposed by Savkoor and Briggs is not
able to capture the physics of sliding adhesive contacts in full. Moreover,
Refs \cite{Vorvolakos2003, Degennes2007} clearly show that a significant
reduction of the contact area only happens at sufficiently high sliding
velocity, and for soft solids always associate with the transition toward a
stick-slip regime. In particular some experiments
\cite{Homola1990,Vorvolakos2003,Degennes2007} show that, at low sliding
velocity, although gross slip condition are already established, no contact
area reduction is observed. It is, then, evident that this behavior cannot be
explained by relying on the Savkoor and Briggs model. Therefore, the contact
area reduction observed in Refs \cite{Homola1990,Degennes2007} as well as in
Fig. 7 in Ref. \cite{Vorvolakos2003} must have a different origin. To this
regard we believe that some coexisting mechanisms can be identified as the
root cause of the contact area reduction (i) the reduction of adhesive bonds
at the interface, (ii) the large non-linear interfacial deformations, and
(iii) the shear stress fluctuations at the interface. The first relies on the
reduction of the adhesive bonds induced by the relative motion between the
contacting surfaces. Indeed, as pointed out firstly by Schallamach
\cite{Schallamach1963} and then by many other authors
\cite{Filippov2004,Vorvolakos2003,Degennes2007,Gravish2010}, this relative
motion leads to an increase of the debonding ratio, whereas the rebinding
ratio remains almost constant. The balance between these ratios can, for
sufficiently high sliding velocity, almost completely mask adhesion. The
second mechanism, causing the observed drop of the contact area below the
Hertz theory prediction, must be ascribed to non-linear large deformations
caused by the tangential tractions at the interface, as indeed reported in
similar range of velocities in Refs. \cite{Degennes2007, Chateauminois2008,
Wu-Bavouzet2010}. The third mechanism, discussed in Sec.
\ref{tangential stress fluctuation}, is related to the genesis of a surface
repulsive energy term associated with the random fluctuations of the shear
stresses at the interface, which is expected to become very significant at the
onset of the stick-slip motion.

It is worth noting that, since the second mechanism is shear dependent, at
very low sliding velocity it can be mitigated by the reduction in shear
stresses at the interface (see Refs \cite{Vorvolakos2003,Chateauminois2008}).
Moreover, for what concerns the first mechanism, referring to Ref
\cite{Schallamach1963}, the adhesive bonds number can be estimated as
$N=N_{0}/\left(  1+V/V_{c}\right)  $, where $N_{0}$ are the number of bonds at
rest, and $V_{c}$ is the critical speed. In the case of PDMS
(Polydimethylsiloxane) adhesive behavior, the latter has been estimated as
$V_{c}\approx10$ mm/s in both Refs \cite{Vorvolakos2003,Degennes2007}.
According to the Schallamach equation, we observe that $N\approx0.9N_{0}$
already at $V\approx1$ mm/s, and it reduces down to $N\approx0.5N_{0}$ at
$V\approx10$ mm/s. We would therefore expect to observe a significant
reduction of the contact area only at velocities in the range $1$-$10$ mm/s,
as indeed observed in Ref. \cite{Vorvolakos2003}. We believe that this
interesting result can be explained by our model by observing that, in the
prescribed range of sliding velocity ($V<1$mm/s), the reduction of adhesive
bonds is negligible, and the JKR static solution is observed even during
sliding. On the contrary, a significant reduction of the contact area compared
to the JKR predictions is observed only for $V>2$-$3$ mm/s (see Fig. 7 in Ref.
\cite{Vorvolakos2003}) when the contact moves from a stable sliding to the
stick-slip regime. In such a case as shown in Sec.
\ref{tangential stress fluctuation} the strong shear stress fluctuations
produce a repulsive surface energy term that hinders the interfacial adhesion.

\section{The effect of tangential stress
fluctuations\label{tangential stress fluctuation}}

So far we have considered the case where the tangential frictional stress is
uniformly distributed at the interface. However, fluctuations in the stress
field may be present, which, as we show below, may have a non-negligible
effect on the contact. In this section we look specifically at this point. So
assume that the tangential frictional stress field at the interface is given
by the sum of an average stress $\tau_{0}$ and a fluctuating term $\tau
_{1}\left(  \mathbf{x}\right)  $, i.e.%
\begin{equation}
\tau\left(  \mathbf{x}\right)  =\tau_{0}+\tau_{1}\left(  \mathbf{x}\right)
\label{non uniform}%
\end{equation}
where the ensemble average $\left\langle \tau_{1}\left(  \mathbf{x}\right)
\right\rangle =0$. The symbol $\left\langle \cdot\right\rangle $ is used in
the sequel to represents the ensemble average operator. Assuming that the
contact area is sufficiently large compared to the wavelengths of the
fluctuating stress field $\tau_{1}\left(  \mathbf{x}\right)  $, local
ergodicity allows us to replace the spatial averages with the ensemble
averages. Now observe that, because of linear elasticity, we can also write
the tangential displacement field $v\left(  \mathbf{x}\right)  $ as the sum of
two terms%
\begin{equation}
v\left(  \mathbf{x}\right)  =w_{0}\left(  \mathbf{x}\right)  +w_{1}\left(
\mathbf{x}\right)  \label{displacementsplitting}%
\end{equation}
where%
\begin{align}
w_{0}\left(  \mathbf{x}\right)   &  =\int d^{2}x^{\prime}G_{x}\left(
\mathbf{x}-\mathbf{x}^{\prime}\right)  \tau_{0}\label{wzero}\\
w_{1}\left(  \mathbf{x}\right)   &  =\int d^{2}x^{\prime}G_{x}\left(
\mathbf{x}-\mathbf{x}^{\prime}\right)  \tau_{1}\left(  \mathbf{x}^{\prime
}\right)  \label{wuno}%
\end{align}
where\ the Green function $G_{x}\left(  \mathbf{x}\right)  $ is given in Eq.
(\ref{equno}). Since $\left\langle w_{1}\left(  \mathbf{x}\right)
\right\rangle =0$, we get $W=\int d^{2}xv\left(  \mathbf{x}\right)  =\int
d^{2}xw_{0}\left(  \mathbf{x}\right)  $. So, the elastic energy of the system
becomes%
\begin{equation}
\mathcal{E}=\frac{1}{2}\int_{\Omega}d^{2}x\sigma\left(  \mathbf{x}\right)
u\left(  \mathbf{x}\right)  +\frac{1}{2}\tau_{0}W+\frac{1}{2}\int_{\Omega
}d^{2}x\tau_{1}\left(  \mathbf{x}\right)  v\left(  \mathbf{x}\right)
\label{elastic energy}%
\end{equation}
Now note that $\left\langle \tau_{1}\left(  \mathbf{x}\right)  v\left(
\mathbf{x}\right)  \right\rangle =\left\langle \tau_{1}\left(  \mathbf{x}%
\right)  w_{1}\left(  \mathbf{x}\right)  \right\rangle $, thus Eq.
(\ref{elastic energy}) can be rephrased as%
\begin{equation}
\mathcal{E}=\frac{1}{2}\int_{\Omega}d^{2}x\sigma\left(  \mathbf{x}\right)
u\left(  \mathbf{x}\right)  +\frac{1}{2}\tau_{0}W+\frac{1}{2}\left\langle
\tau_{1}\left(  \mathbf{x}\right)  w_{1}\left(  \mathbf{x}\right)
\right\rangle A \label{final elastic energy}%
\end{equation}
where because of ergodicity the quantity $\frac{1}{2}\left\langle \tau
_{1}\left(  \mathbf{x}\right)  w_{1}\left(  \mathbf{x}\right)  \right\rangle
=\Delta\Gamma$ is independent of the position vector $\mathbf{x}$\ and
represents a surface energy per unit area. Using Eq. (\ref{wuno}) this surface
energy $\Delta\Gamma$ can be easily calculated as
\begin{equation}
\left\langle \tau_{1}\left(  \mathbf{x}\right)  w_{1}\left(  \mathbf{x}%
\right)  \right\rangle =\int d^{2}x^{\prime}G_{x}\left(  \mathbf{x}%
-\mathbf{x}^{\prime}\right)  \left\langle \tau_{1}\left(  \mathbf{x}\right)
\tau_{1}\left(  \mathbf{x}^{\prime}\right)  \right\rangle =\int d^{2}%
xG_{x}\left(  \mathbf{x}\right)  \left\langle \tau_{1}\left(  \mathbf{x}%
\right)  \tau_{1}\left(  \mathbf{0}\right)  \right\rangle
\label{in real space}%
\end{equation}
Eq. (\ref{in real space}) has also an analogous expression in Fourier space
which may be easier to calculate. Note that by taking the Fourier transform of
Eq. (\ref{equno}) we get%
\begin{equation}
G_{x}\left(  \mathbf{q}\right)  =\int d^{2}xG_{x}\left(  \mathbf{x}\right)
e^{-i\mathbf{q\cdot x}}=\frac{1}{2G}\frac{1}{q}\left(  1+\frac{q_{y}^{2}%
}{q^{2}}\right)  \label{tangetialcomplex modulus}%
\end{equation}
then taking the Fourier transform of Eq. (\ref{wuno}) we also obtain%
\begin{equation}
w_{1}\left(  \mathbf{q}\right)  =G_{x}\left(  \mathbf{q}\right)  \tau
_{1}\left(  \mathbf{q}\right)  . \label{elasticfourier}%
\end{equation}
Using%
\begin{equation}
\left\langle \tau_{1}\left(  \mathbf{x}\right)  w_{1}\left(  \mathbf{x}%
\right)  \right\rangle =\frac{1}{\left(  2\pi\right)  ^{4}}\int d^{2}%
qd^{2}q^{\prime}\left\langle \tau_{1}\left(  \mathbf{q}\right)  w_{1}\left(
\mathbf{q}^{\prime}\right)  \right\rangle e^{i\mathbf{q\cdot x}}%
e^{i\mathbf{q}^{\prime}\mathbf{\cdot x}} \label{calculus1}%
\end{equation}
and replacing Eq. (\ref{elasticfourier}) in Eq. (\ref{calculus1}) we get%
\begin{equation}
\left\langle \tau_{1}\left(  \mathbf{x}\right)  w_{1}\left(  \mathbf{x}%
\right)  \right\rangle =\frac{1}{\left(  2\pi\right)  ^{4}}\int d^{2}%
qd^{2}q^{\prime}G_{x}\left(  \mathbf{q}^{\prime}\right)  \left\langle \tau
_{1}\left(  \mathbf{q}\right)  \tau_{1}\left(  \mathbf{q}^{\prime}\right)
\right\rangle e^{i\mathbf{q\cdot x}}e^{i\mathbf{q}^{\prime}\mathbf{\cdot x}}
\label{crosscorrelation}%
\end{equation}
The quantity $\left\langle \tau_{1}\left(  \mathbf{q}\right)  \tau_{1}\left(
\mathbf{q}^{\prime}\right)  \right\rangle =\left(  2\pi\right)  ^{2}C_{\tau
}\left(  \mathbf{q}\right)  \delta\left(  \mathbf{\mathbf{q+}q}^{\prime
}\right)  $, where the power spectral density of the stress fluctuation at the
interface is $C_{\tau}\left(  \mathbf{q}\right)  =\int d^{2}x\left\langle
\tau\left(  \mathbf{x}\right)  \tau\left(  \mathbf{0}\right)  \right\rangle
e^{-i\mathbf{q\cdot x}}$. Therefore we obtain%
\begin{equation}
\left\langle \tau_{1}\left(  \mathbf{x}\right)  w_{1}\left(  \mathbf{x}%
\right)  \right\rangle =\int d^{2}xG_{x}\left(  \mathbf{x}\right)
\left\langle \tau\left(  \mathbf{x}\right)  \tau\left(  \mathbf{0}\right)
\right\rangle =\frac{1}{\left(  2\pi\right)  ^{2}}\int d^{2}qG_{x}\left(
\mathbf{q}\right)  C_{\tau}\left(  \mathbf{q}\right)  \label{surface1}%
\end{equation}
which allows us to calculate the surface energy per unit area $\Delta\Gamma$
as
\begin{equation}
\Delta\Gamma=\frac{1}{8\pi^{2}}\int d^{2}qG_{x}\left(  \mathbf{q}\right)
C_{\tau}\left(  \mathbf{q}\right)  >0 \label{surface2}%
\end{equation}
and the elastic energy%
\begin{equation}
\mathcal{E}=\frac{1}{2}\int_{\Omega}d^{2}x\sigma\left(  \mathbf{x}\right)
u\left(  \mathbf{x}\right)  +\frac{1}{2}\tau_{0}W+\Delta\Gamma A=\mathcal{E}%
_{0}+\Delta\Gamma A. \label{elastic energy fluctuation}%
\end{equation}
The total energy of the system is then%
\begin{equation}
\mathcal{U}\left(  s,W,A\right)  =\mathcal{E}_{0}\left(  s,W,A\right)
-\left(  \Delta\gamma-\Delta\Gamma\right)  A.
\label{total internal energy fluctuation}%
\end{equation}
Equation (\ref{total internal energy fluctuation}) shows that the effect of
tangential stress fluctuation $\tau_{1}\left(  \mathbf{x}\right)  $ at the
interface consists in including a `repulsive' surface energy term per unit
area, which diminishes the adhesion energy $\Delta\gamma$ per unit area of the
amount $\Delta\Gamma$. Thus, by replacing $\Delta\gamma\rightarrow\Delta
\gamma-\Delta\Gamma$ the original formulation of Sec. \ref{formulation}, is
easily retrieved. Note that if the fluctuations of the interfacial tangential
stresses are sufficiently large [i.e. significant values of $C_{\tau}\left(
\mathbf{q}\right)  $] the term $\Delta\Gamma$ may become comparable with
$\Delta\gamma$ and, in the end, may even counterbalance or overcome the effect
of the uniform stress at the interface, thus leading to a reduction of the
contact area. Indeed, such fluctuations are strongly enhanced at the onset of
stick-slip motion. This may explain why above a certain velocity threshold,
$V>2$-$3$ mm/s, a strong reduction of contact area is observed, as shown in
Ref. \cite{Vorvolakos2003}.

To estimate the amplitude of shear stress fluctuations necessary to completely
mask the adhesion energy, let us now consider the simple case of a sinusoidal
fluctuation of the shear stress field, i.e. let us assume that $\tau
_{1}\left(  \mathbf{x}\right)  =\tau_{1}\cos\left(  q_{k}x+\varphi\right)
=\frac{1}{2}\tau_{1}\left[  \exp\left(  iq_{k}x\right)  \exp\left(
i\varphi\right)  +\exp\left(  -iq_{k}x\right)  \exp\left(  -i\varphi\right)
\right]  $, where $0\leq\varphi<2\pi$ is a uniformly distributed random phase.
Then we obtain
\begin{equation}
\left\langle \tau\left(  \mathbf{x}\right)  \tau\left(  \mathbf{0}\right)
\right\rangle =\frac{1}{4}\tau_{1}^{2}\left(  e^{iq_{k}x}+e^{-iq_{k}x}\right)
=\frac{1}{2}\tau_{1}^{2}\cos\left(  q_{k}x\right)  \label{stress correlation}%
\end{equation}
Taking the Fourier transform we obtain%
\begin{equation}
C_{\tau}\left(  \mathbf{q}\right)  =\int d^{2}x\left\langle \tau\left(
\mathbf{x}\right)  \tau\left(  \mathbf{0}\right)  \right\rangle
e^{-i\mathbf{q\cdot x}}=\pi^{2}\tau_{1}^{2}\delta\left(  q_{y}\right)
\left\{  \delta\left(  q_{x}-q_{k}\right)  +\delta\left(  q_{x}+q_{k}\right)
\right\}  \label{PSD sinus}%
\end{equation}
substituting Eq. (\ref{PSD sinus}) into Eq. (\ref{surface2}) we get%
\begin{equation}
\Delta\Gamma=\frac{\tau_{1}^{2}}{8G}\frac{1}{q_{k}}=\frac{\tau_{1}^{2}}{16\pi
G}\lambda_{k}=\frac{\tau_{1}^{2}}{16\pi G}\frac{L}{k}%
\end{equation}
where $L\approx2a$ is the lateral size of the contact area. Therefore the
amplitude $\tau_{1}$ of the fluctuating stress needed to fully cancel the
effect of adhesion, i.e. to make $\Delta\Gamma=\Delta\gamma$ is%
\[
\tau_{1}=4\sqrt{\pi k\frac{G\Delta\gamma}{L}}.
\]
Assuming $k=10$, and as in Ref. \cite{Vorvolakos2003}, $G=1.6\mathrm{MPa}$,
$L=0.24\mathrm{mm}$, $\Delta\gamma=42\mathrm{mJ/m}^{2}$ we get $\tau
_{1}\approx375\mathrm{kPa}$. Using the data from Ref. \cite{Chateauminois2008}%
, i.e. $G=0.5\mathrm{MPa}$, $L=4\mathrm{mm}$, and $\Delta\gamma
=42\mathrm{mJ/m}^{2}$ we get $\tau_{1}\approx51\mathrm{kPa}$. Therefore, the
intensity of shear stress fluctuations needed to completely mask adhesion is
as large as the average stress $\tau_{0}$ at the interface. Such large
fluctuation may occur in the contact area of soft-solids at the one-set of
stick-slip. Should these stress-fluctuation really be observed under
stick-slip conditions, they would provide an additional important step toward
the complete understanding of the physics governing the reduction of the size
of the contact area in sliding contacts. This emphasizes the need for further
experimental investigations in which fluctuations, as well as contact area and
frictional force, are captured to shed light on the complex interplay between
adhesion and tangential interfacial stresses.

\section{Conclusions}

In this work we have investigated the effect of interfacial tangential
tractions on the contact area evolution in adhesive sliding contacts, under
gross slip conditions. We developed a theoretical model which, relying on
energetic arguments, takes into account the mechanical energy term related to
the applied tangential tractions, regardless of their nature and cause
(frictional, chemical, etc.). We focused on the exemplar case of a rigid
smooth sphere in sliding contact with an elastic half-space.

The model shows that \textit{the presence of uniform tangential stresses at
the interface does not modify the effective energy of adhesion, therefore the
contact behavior follows the classical JKR\ solution even in sliding at
uniform frictional stress.} This result is in agreement with the experiments
carried out by Vorvolakos and Chaudhury in 2003 (Ref. \cite{Vorvolakos2003}),
who observed experimentally that the contact area of a PDMS\ sphere remains
constant during sliding, at least up to velocity of 1 mm/s, and is in
agreement with the JKR solution. Our result also explain the experimental
outcomes of Carpick et al. in Ref. \cite{Carpick}, who observed that the
friction force between a platinum-coated atomic force microscope (AFM) tip and
the surface of mica in ultrahigh vacuum (UHV) varies with load in proportion
to the contact area predicted by the Johnson-Kendall-Roberts (JKR).

A reduction of contact area at higher sliding velocity may originate from a
reduction of the density of adhesive bonds as the velocity is increased, or
caused by the repulsive energy term $\Delta\Gamma=\frac{1}{2}\left\langle
\tau_{1}\left(  \mathbf{x}\right)  w_{1}\left(  \mathbf{x}\right)
\right\rangle $ associated with the stress spatial fluctuation at the
interface. This may explain why adhesion is completely masked at relatively
large sliding velocities. This scenario seems to be, at least partially,
confirmed in the scientific experimental literature.

\begin{acknowledgement}
The authors thanks Bo Persson, Michele Ciavarella, and Antoine Chateauminois
for valuable scientific discussions, as well as the referees for the
interesting suggestions. D.D. would like to acknowledge the support received
from the Engineering and Physical Sciences Research Council (EPSRC) via his
Established Career Fellowship EP/N025954/1.
\end{acknowledgement}

\appendix

\section{The elastic tangential displacement field over a circular region
loaded with uniformly distributed tangential stresses.\label{appendice1}}

In this section we calculate the tangential displacement field due to uniform
and unidirectional tangential traction acting along the $x$ axis on an elastic
half-space surface and distributed over circular area of radius $a$. We focus,
for simplicity, on incompressible materials (so that there is no interaction
between normal and tangential fields) with Poisson's ratio $\nu=0.5$. In such
a case, we recall that the surface tangential displacement $G_{x}\left(
\mathbf{x}\right)  $ due to a concentrated unit force $F_{x}=1$ placed at the
origin of the half-space and tangentially directed along the $x$ axis is (see
Refs \cite{landau,Johnson})%
\begin{equation}
G_{x}\left(  \mathbf{x}\right)  =\frac{1+\nu}{2\pi E}\frac{1}{\left\vert
\mathbf{x}\right\vert }\left[  2\left(  1-\nu\right)  +2\nu\frac{x^{2}%
}{\left\vert \mathbf{x}\right\vert ^{2}}\right]  =\frac{1}{4\pi G}\frac
{1}{\left\vert \mathbf{x}\right\vert }\left(  1+\frac{x^{2}}{\left\vert
\mathbf{x}\right\vert ^{2}}\right)  =\frac{1}{4\pi G}\frac{1}{\left\vert
\mathbf{x}\right\vert }+\frac{1}{4\pi G}\frac{x^{2}}{\left\vert \mathbf{x}%
\right\vert ^{3}} \label{equno}%
\end{equation}
where $G$ is the shear modulus.

The surface tangential displacement due to a uniform tangential stress
$\tau_{x}\left(  \mathbf{x}\right)  =\tau_{0}$ acting on a circular area can
be found as the convolution integral on the contact area of Eq. \ref{equno}%
\begin{equation}
v\left(  \mathbf{x}\right)  =\frac{\tau_{0}}{4\pi G}\int d^{2}x^{\prime}%
\frac{1}{\left\vert \mathbf{x}-\mathbf{x}^{\prime}\right\vert }+\frac{\tau
_{0}}{4\pi G}\int d^{2}x^{\prime}\frac{\left(  x-x^{\prime}\right)  ^{2}%
}{\left\vert \mathbf{x}-\mathbf{x}^{\prime}\right\vert ^{3}}=v_{1}\left(
\mathbf{x}\right)  +v_{2}\left(  \mathbf{x}\right)  \label{eqdue}%
\end{equation}
The first term in Eq. (\ref{eqdue}) is the analogues of the normal
displacement at the interface obtained with the application of a uniform
pressure distribution on a circular area of radius $a$. The solution is given
by Johnson. In such a case we get that within the loaded circle
\begin{equation}
v_{1}\left(  \mathbf{x}\right)  =\frac{\tau_{0}}{4\pi G}\int d^{2}x^{\prime
}\frac{1}{\left\vert \mathbf{x}-\mathbf{x}^{\prime}\right\vert }=\frac
{\tau_{0}a}{\pi G}E\left(  \rho\right)  ;\qquad\rho<1 \label{v1}%
\end{equation}
where $\rho=r/a$, and $E\left(  \rho\right)  =\int_{0}^{\pi/2}\sqrt{1-\rho
^{2}\sin^{2}\varphi}d\varphi$ is the elliptic integral of the second kind.

Furthermore, by defining $\mathbf{s}=\mathbf{x}-\mathbf{x}^{\prime}$ and
\begin{align}
x^{\prime}-x  &  =s\cos\varphi\label{mapping}\\
y^{\prime}-y  &  =s\sin\varphi
\end{align}
the second term in Eq. (\ref{eqdue}) can be rewritten as%
\begin{equation}
v_{2}\left(  \mathbf{x}\right)  =\frac{\tau_{0}}{4\pi G}\int_{0}^{2\pi
}d\varphi\cos^{2}\varphi\int_{0}^{s_{1}}ds=\frac{\tau_{0}}{4\pi G}\int
_{0}^{2\pi}d\varphi\sqrt{a^{2}-r^{2}\sin^{2}\left(  \varphi-\theta\right)
}\cos^{2}\varphi\label{v2_uno}%
\end{equation}
where $s_{1}\left(  \varphi\right)  =-x\cos\varphi-y\sin\varphi+\sqrt
{a^{2}-\left(  x\sin\varphi-y\cos\varphi\right)  ^{2}}$, $x=r\cos\theta$,and
$y=r\sin\theta.$

Finally, after a few algebraic manipulations Eq. (\ref{v2_uno}) gives%
\begin{equation}
v_{2}\left(  \mathbf{x}\right)  =\frac{\tau_{0}a}{2\pi G}\left[  1+\frac{1}%
{3}\frac{2-\rho^{2}}{\rho^{2}}\cos\left(  2\theta\right)  \right]  E\left(
\rho\right)  -\frac{1}{3}\frac{\tau_{0}a}{\pi G}\frac{1-\rho^{2}}{\rho^{2}%
}\cos\left(  2\theta\right)  K\left(  \rho\right)  ;\qquad\rho<1
\label{v2_due}%
\end{equation}
where $K\left(  \rho\right)  =\int_{0}^{\pi/2}d\varphi\left(  1-\rho^{2}%
\sin^{2}\varphi\right)  ^{-1/2}$ is complete Elliptic integral of first kind

Therefore, combining Eqs. (\ref{v1},\ref{v2_due}) with Eq. (\ref{eqdue}), the
tangential displacement field due to uniform and unidirectional tangential
tractions over a circular contact is%
\begin{align}
v\left(  \mathbf{x}\right)   &  =v_{1}\left(  \mathbf{x}\right)  +v_{2}\left(
\mathbf{x}\right) \label{A8}\\
&  =\frac{3}{2}\frac{\tau_{0}a}{\pi G}E\left(  \rho\right)  +\frac{1}{3}%
\frac{\tau_{0}a}{2\pi G}\frac{2-\rho^{2}}{\rho^{2}}E\left(  \rho\right)
\cos\left(  2\theta\right)  -\frac{1}{3}\frac{\tau_{0}a}{\pi G}\frac
{1-\rho^{2}}{\rho^{2}}K\left(  \rho\right)  \cos\left(  2\theta\right)
;\qquad\rho<1\nonumber
\end{align}

Notably, the mean displacement $v_{m}$ in the contact area is
\begin{equation}
v_{m}=\frac{1}{\pi a^{2}}\int v\left(  \mathbf{x}\right)  d^{2}x=\frac{1}{\pi
a^{2}}\int drd\theta rv\left(  r,\theta\right)  =\frac{2\tau_{0}a}{\pi G}
\label{A9}%
\end{equation}

\end{document}